\documentclass[aps,pra,twocolumn,showpacs]{revtex4-1}
\usepackage{epsfig}
\usepackage{bm}

\newcommand{\be}{\begin{eqnarray}}
\newcommand{\ee}{\end{eqnarray}}
\newcommand{\ba}{\begin{array}}
\newcommand{\ea}{\end{array}}
\newcommand{\bmat}{\left(\begin{array}}
\newcommand{\emat}{\end{array}\right)}
\newcommand{\no}{\nonumber}
\newcommand{\diff}{\mathrm d}
\newcommand{\e}{\mathrm e}

\begin{document}
\title{Fast-forward scaling in a finite-dimensional Hilbert space}
\author{Kazutaka Takahashi}
\affiliation{Department of Physics, Tokyo Institute of Technology, 
Tokyo 152-8551, Japan}

\date{\today}

\begin{abstract}
Time evolution of quantum systems is accelerated by the fast-forward scaling.
We reformulate the method to study systems 
in a finite-dimensional Hilbert space.
For several simple systems, 
we explicitly construct the acceleration potential.
We also use our formulation to accelerate the adiabatic dynamics.
Applying the method to the transitionless quantum driving,
we find that the fast-forward potential can be understood 
as a counterdiabatic term.
\end{abstract}
\maketitle

%%%%%%%%%%%%%%%%%%%%%%%%%%%%%%%%%%%%%%%%%%%%%%%%%%%%%%%%%%%%%%%%%%%%%%%%%%%
\section{Introduction}

Recent developments in techniques of controlling quantum systems 
have brought about inventing new theoretical methods of acceleration.
In the methods called 
the assisted adiabatic passage~\cite{DR1,DR2}
or the transitionless quantum driving~\cite{Berry}, 
the acceleration of the adiabatic state is realized by applying 
the counterdiabatic Hamiltonian.
The method is known to be equivalent to 
the Lewis-Riesenfeld invariant-based engineering~\cite{LR,CRSCGM}.
The optimal driving Hamiltonian is constructed 
under the condition that there exists an invariant quantity 
throughout the time evolution.
Although the formulation is different between their methods, 
the essential mechanism is shown to be the same. 
We therefore call them generally  
``shortcuts to adiabaticity''~\cite{STA}.
Furthermore, these methods are derived from 
the quantum brachistochrone equation 
by imposing a proper constraint~\cite{CHKO,Takahashi},
which shows that they admit a unified interpretation.
The method has been implemented experimentally~\cite{SSVL,SSCVL,Betal,Zetal}
and is important also for practical applications. 

The fast-forward scaling method 
proposed by Masuda and Nakamura
is known to be different from the other methods 
to design shortcuts to adiabaticity and 
plays a unique role in the methods of acceleration~\cite{MN1,MN2,MN3,TMRM}.
The method is formulated in the coordinate representation of 
the Schr\"odinger equation.
It is applied to an evolving wave packet, 
either in noninteracting matter wave of a Bose-Einstein condensate
described by the Gross-Pitaevski equation.
The main idea is to determine the auxiliary accelerating Hamiltonian by 
a local potential term.
To find the acceleration potential, the unitary transformation 
is performed for the original state.

From a more general perspective, 
we expect that there exists some relation between the fast-forward 
scaling method and the other techniques to design 
shortcuts to adiabaticity, mentioned above.
In the fast-forward scaling, 
the form of the acceleration potential depends explicitly on 
the wavefunction to accelerate and takes a complicated form.
The original formulation of the fast-forward method is limited 
to matter waves described as continuous variable systems. 
The extension of this method to finite-dimensional Hilbert space 
remains an interesting open problem.
In Ref.~\cite{MR}, Masuda and Rice formulated 
the fast-forward scaling in lattice systems 
aiming at the application to 
a Bose-Einstein condensate in an optical lattice.
This formulation can be more generalized 
to treat other discrete systems such as spin models.

In this paper, we formulate the method in an arbitrary
finite-dimensional Hilbert space.
Although the main calculation treats two-level systems,
it is a straightforward task to extend the formulation to 
higher-dimensional systems.
Furthermore, we study how the method is related to 
the acceleration of the adiabatic state.
Applying the fast-forward scaling to 
the formula of the transitionless driving, 
we show that the fast-forwarded state is also transitionless. 

%%%%%%%%%%%%%%%%%%%%%%%%%%%%%%%%%%%%%%%%%%%%%%%%%%%%%%%%%%%%%%%%%%%%%%%%%%%
\section{Fast-forward scaling}

%%%%%%%%%%%%%%%%%%%%%%%%%%%%%%%%%%%%%%%%%%%%%%%%%%%%%%%%%%%%%%%%%%%%%%%%%%%
\subsection{Formulation}

We start from the time-dependent Schr\"odinger equation 
with the Hamiltonian $\hat{H}(t)$
\be
 i\frac{\diff}{\diff t}|\psi(t)\rangle = \hat{H}(t)|\psi(t)\rangle.
\ee
For a given Hamiltonian, we assume that 
we have a solution of state $|\psi(t)\rangle$ in hand.
We want to accelerate this time evolution 
by applying an external potential.
In the fast-forward scaling, the time $t$ is reparametrized as 
\be
 \Lambda(t) = \int_0^t \diff t'\,\alpha(t'),
\ee
where $\alpha(t)$ is real and is greater than or equal to unity.
In this new scale, the evolution of the state is fast-forwarded.
Then, for a reason described below, we apply a unitary transformation
\be
 \hat{U}(t)= \e^{-i\hat{f}(t)}, \label{unitary2}
\ee
where $\hat{f}(t)$ is a Hermitian operator.
The Schr\"odinger equation is rewritten as 
\be
 & & i\frac{\diff}{\diff t}|\psi_{\rm FF}(t)\rangle 
 = \hat{H}_{\rm FF}(t)|\psi_{\rm FF}(t)\rangle,
\ee
 where 
\be
 & & |\psi_{\rm FF}(t)\rangle = \hat{U}(t)|\psi(\Lambda(t))\rangle, \\
 & & \hat{H}_{\rm FF}(t)= 
 \hat{U}(t)\left(-i\frac{\diff}{\diff t}\hat{U}^\dag(t)\right)
 +\alpha(t)\hat{U}(t)\hat{H}(\Lambda(t))\hat{U}^\dag(t). \no\\
\ee

The idea of the fast-forward scaling is to write 
the fast-forward Hamiltonian as 
the sum of the original Hamiltonian and 
the acceleration potential:
\be
 \hat{H}_{\rm FF}(t) \sim \hat{H}(t)+\hat{V}(t).
\ee
The operators in the left- and right-hand sides are not equal to each other.
The relation with the symbol $\sim$ means that 
the operation to the state $|\psi_{\rm FF}(t)\rangle$ gives the same effect as 
\be
 \hat{H}_{\rm FF}(t)|\psi_{\rm FF}(t)\rangle
 =(\hat{H}(t)+\hat{V}(t))|\psi_{\rm FF}(t)\rangle.
\ee
In the original study, the acceleration potential $\hat{V}(t)$ 
is represented by a local potential term.
It can be possible by choosing the unitary transformation in a proper way.

%%%%%%%%%%%

We apply the method to discrete systems such as spin systems.
As the simplest case, we use a two-level Hamiltonian
\be
 \hat{H}(t) &=& \frac{1}{2}\bm{h}(t)\cdot\bm{\sigma} \no\\
 &=& \frac{1}{2}
 \bmat{cc} h_3(t) & h_1(t)-ih_2(t) \\ h_1(t)+ih_2(t) & -h_3(t) \emat,
\ee
where $\bm{h}(t)=(h_1(t),h_2(t),h_3(t))$ 
is a three-dimensional magnetic-field vector and 
each component of $\bm{\sigma}=(\sigma_x,\sigma_y,\sigma_z)$ 
denotes a Pauli matrix.

First, we discuss how to choose the operator $\hat{f}$ 
in Eq.~(\ref{unitary2}).
In the original analysis, 
$f$ is a coordinate-dependent operator as $\hat{f}=f(\hat{x},t)$.
In the present two-level systems, 
anticipating that we measure the spin in $z$ direction, 
we choose
\be
 \hat{f}(t) = \frac{\phi(t)}{2}\sigma_z,\label{f}
\ee
where $\phi(t)$ is a real scalar function determined below.
The probability that the state $|\psi(t)\rangle$ is observed in  
the up or down spin state $|\sigma=\pm 1\rangle$ is unchanged 
under the unitary transformation:
\be
 |\langle \sigma|\psi(t)\rangle|^2
 =|\langle \sigma|\e^{-i\phi(t)\sigma_z/2}|\psi(t)\rangle|^2.
\ee

We can also consider an operator $\hat{f}(t)$ 
which is proportional to the identity operator.
It does not give any quantum effects and is not important.
By using the gauge transformation,
we can eliminate the identity-operator term in the acceleration potential
as we mention below.

Second, under the choice of $\hat{f}(t)$ in Eq.~(\ref{f}), 
we separate the Hamiltonian into two parts: 
\be
 & & \hat{H}_{\rm FF}(t)
 = \frac{1}{2}\Bigl(\bm{h}(t)\cdot\bm{\sigma}
 +\tilde{\bm{h}}(t)\cdot\bm{\sigma}\Bigr), \\
 & & \tilde{\bm{h}}(t)=\bmat{c} 
 \alpha\Bigl(h_{1\Lambda}\cos\phi
 -h_{2\Lambda}\sin\phi\Bigr)-h_1 \\
 \alpha \Bigl(h_{1\Lambda}\sin\phi
 +h_{2\Lambda}\cos\phi\Bigr)-h_2 \\
 \dot{\phi}+\alpha h_{3\Lambda}-h_3 \emat,
\ee
where $h_{1\Lambda}=h_1(\Lambda(t))$
and $\dot{\phi}$ is the time derivative of $\phi$.
The first term of $\hat{H}_{\rm FF}(t)$ is 
the original Hamiltonian before the scaling.
To write the second term by using $\sigma_z$ only, 
we need to know the explicit form of the wavefunction.
We write the original state as 
\be
 & & |\psi(t)\rangle = \bmat{c} a(t) \\ b(t) \emat 
 = a(t)|+\rangle +b(t)|-\rangle.
\ee
The ``coordinate'' representation of the state 
$\psi_{\rm FF}(\sigma,t)=\langle\sigma|\psi_{\rm FF}(t)\rangle$ 
is 
\be
 \psi_{\rm FF}(\sigma,t)
 &=& \left\{\ba{cc} \e^{-i\phi/2}a_\Lambda \\ \e^{i\phi/2}b_\Lambda \ea\right\} \no\\
 &=& \e^{-i\phi/2}a_\Lambda\frac{1+\sigma}{2}
 +\e^{i\phi/2}b_\Lambda\frac{1-\sigma}{2}, 
\ee
where $a_{\Lambda}=a(\Lambda(t))$ and $b_{\Lambda}=b(\Lambda(t))$.
The symbol $\sigma$ takes $\pm 1$, which plays the role of ``$x$''.
In the same way, we can write 
\be
\langle\sigma|\sigma_z|\psi_{\rm FF}\rangle 
 &=& \left\{\ba{cc} \e^{-i\phi/2}a_\Lambda \\ -\e^{i\phi/2}b_\Lambda \ea\right\} \no\\
 &=& \sigma\psi_{\rm FF}, \\
 \langle\sigma|\sigma_x|\psi_{\rm FF}\rangle 
 &=& \left\{\ba{cc} \e^{i\phi/2}b_\Lambda \\ \e^{-i\phi/2}a_\Lambda \ea\right\} \no\\
 &=& \left(\frac{b_\Lambda}{a_\Lambda}\e^{i\phi}\frac{1+\sigma}{2}
 +\frac{a_\Lambda}{b_\Lambda}\e^{-i\phi}\frac{1-\sigma}{2}\right)
 \psi_{\rm FF}, \\
 \langle\sigma|\sigma_y|\psi_{\rm FF}\rangle 
 &=& \left\{\ba{cc} -i\e^{i\phi/2}b_\Lambda \\ i\e^{-i\phi/2}a_\Lambda \ea\right\} \no\\
 &=& -i\left(\frac{b_\Lambda}{a_\Lambda}\e^{i\phi}\frac{1+\sigma}{2}
 -\frac{a_\Lambda}{b_\Lambda}\e^{-i\phi}\frac{1-\sigma}{2}\right)\psi_{\rm FF}. \no\\
\ee
Using these relations, we obtain the acceleration potential 
$\hat{V}(t)={\rm diag}(V(+1,t),V(-1,t))$ where 
\be
 V(\sigma,t) &=& \frac{1}{2}
 (\tilde{h}_1-i\tilde{h}_2)
 \frac{b_\Lambda}{a_\Lambda}\e^{i\phi}\frac{1+\sigma}{2} \no\\
 && +\frac{1}{2}(\tilde{h}_1+i\tilde{h}_2)
 \frac{a_\Lambda}{b_\Lambda}\e^{-i\phi}\frac{1-\sigma}{2} 
 +\frac{1}{2}\tilde{h}_3\sigma. \label{Vff}
\ee
The value of $\phi(t)$ is determined so that the potential is real.
The condition is given by 
\be
% \tilde{h}_1(t){\rm Im}\left(\tilde{a}^*(t)\tilde{b}(t)\right)
% =\tilde{h}_2(t){\rm Re}\left(\tilde{a}^*(t)\tilde{b}(t)\right)
 (\tilde{h}_1-i\tilde{h}_2)
 \frac{b_\Lambda}{a_\Lambda}\e^{i\phi}
 =(\tilde{h}_1+i\tilde{h}_2)
 \frac{b_\Lambda^*}{a_\Lambda^*}\e^{-i\phi}.
 \label{cond1}
\ee
Thus the acceleration potential has the form 
$V(\sigma,t)=(v_0(t)+v(t)\sigma)/2$ with 
real $v_0(t)$ and $v(t)$.
Since the term $v_0(t)$ only affects the overall phase of the state,
it does not play any role for the acceleration.
We can eliminate this term by using a unitary transformation
(\ref{unitary2}) with the form $\hat{f}(t)=(\phi_0(t)+\phi(t)\sigma_z)/2$.

%%%%%%%%%%%%%%%%%%%%%%%%%%%%%%%%%%%%%%%%%%%%%%%%%%%%%%%%%%%%%%%%%%%%%%%%%%%
\subsection{Example: two-level system}
\label{ex}

To see how the method works, 
we treat an example of a two-level system.
The magnetic field is chosen to be 
\be
 \bm{h}(t)=\bmat{c}
 h_0\cos\omega t \\ -\omega \\ h_0\sin\omega t
 \emat.
\ee
For simplicity, we set $h_0=1$ and 
make all variables dimensionless in the following calculations. 
The Hamiltonian reads 
\be
 \hat{H}(t)=\frac{1}{2}\bmat{cc} \sin\omega t & \cos\omega t+i\omega \\ 
 \cos\omega t-i\omega & -\sin\omega t \emat.
\ee
One of the exact solutions of the Schr\"odinger equation is given by 
\be
 |\psi(t)\rangle = \frac{1}{\sqrt{2}}\e^{-it/2}\bmat{c}
 \cos\frac{\omega t}{2}+\sin\frac{\omega t}{2} \\
 \cos\frac{\omega t}{2}-\sin\frac{\omega t}{2} 
 \emat.
\ee
We set the initial state at $t=0$ 
is given by the eigenstate of $\sigma_x$ with the eigenvalue $+1$.
With the time evolution, the spin 
rotates around the  $y$ axis and 
points to the positive-$z$ direction 
at $t=t_{\rm f}=\pi/2\omega$.
We are interested in fast-forwarding this motion.

The condition (\ref{cond1}) is explicitly written as 
\be
 \omega\alpha(t)-\cos\omega t\sin\phi(t)
 = \omega \cos\phi(t). \label{condex}
\ee
Using the obtained function $\phi(t)$,
we can write the potential
\be
 V(\sigma,t) &=& \frac{1}{2}\left(\alpha
 +\frac{\omega\sin\phi-\cos\phi\cos\omega t}{\cos\omega\Lambda}\right) \no\\
 && 
 \no\\
 && +\frac{1}{2}\biggl[\dot{\phi}
 -\frac{(\omega\sin\phi-\cos\phi\cos\omega t)\sin\omega\Lambda}{\cos\omega\Lambda}
 \no\\
 && \qquad
 -\sin\omega t\biggr]\sigma.
 \label{Vs}
\ee

\begin{center}
\begin{figure}[t]
\begin{center}
\includegraphics[width=0.9\columnwidth]{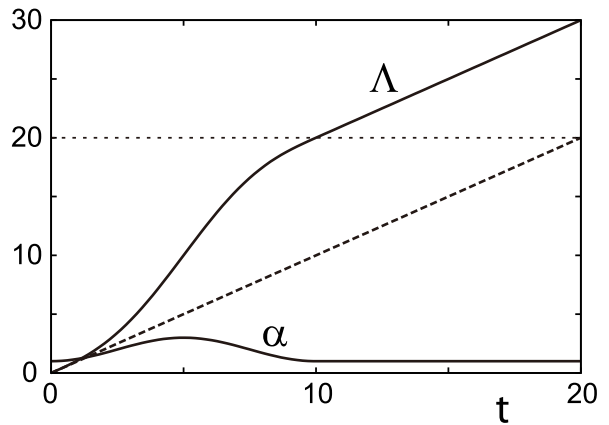}
\end{center}
\caption{
Protocols (\ref{alpha}) and (\ref{lambda}) at 
$\bar{\alpha}=2.0$ and $t_0=10$.
The dashed line represents the time before the scaling.}
\label{fig1}
\end{figure}
\end{center}

\begin{center}
\begin{figure}[t]
\begin{center}
\includegraphics[width=0.9\columnwidth]{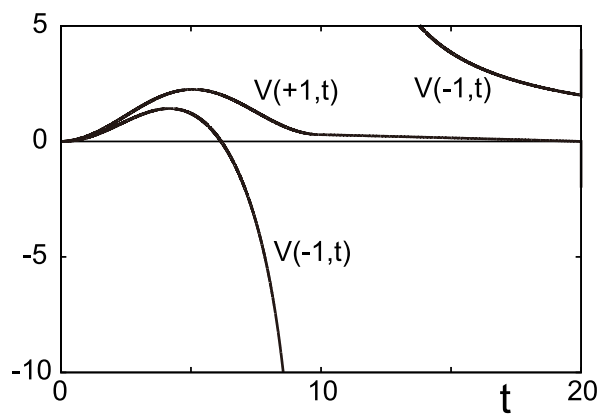}
\end{center}
\caption{
Acceleration potential $V(\sigma=\pm 1,t)$ in Eq.~(\ref{Vs}).
$V(-1,t)$ at $t=10$ goes to $-\infty$ from the left 
and $\infty$ from the right.
}
\label{fig2}
\end{figure}
\end{center}

Following the original analysis~\cite{MN1}, 
we choose the magnification factor $\alpha(t)$ as 
\be
 \alpha(t)=
 \left\{\ba{cc}
 \bar{\alpha}+
 (1-\bar{\alpha})\cos\left(\frac{2\pi t}{t_{\rm 0}}\right) & 0\le t\le t_0 \\
 1 & t>t_0 \\
 \ea\right., \label{alpha}
\ee
where $\bar{\alpha}>1$.
$\Lambda(t)$ is given by 
\be
 \Lambda(t)=
 \left\{\ba{cc}
 \bar{\alpha}t
 +(1-\bar{\alpha})\frac{\sin\left(\frac{2\pi t}{t_0}\right)}
 {\frac{2\pi}{t_0}} & 0\le t\le t_0 \\
 \bar{\alpha}t_0+t-t_0 & t>t_0 \\
 \ea\right..
 \label{lambda}
\ee
These functions are plotted in Fig.~\ref{fig1} 
for $\bar{\alpha}=2$ and $t_0=10$.
The final time $t_{\rm f}=\pi/2\omega=20$ before the fast-forwarding 
is shortened as $10$ by the scaling.
Correspondingly, the behavior of the acceleration potential $V(\sigma,t)$
is plotted as in Fig.~\ref{fig2}.
We see that the potential $V(\sigma=-1,t)$ diverges at the final time 
$t=10$.
This behavior is understood in the general expression 
of the potential in Eq.~(\ref{Vff}) where one of the components of 
the state $a_\Lambda$ in the denominator goes to zero.

%%%%%%%%%%%%%%
%%%%%%%%%%%%
%%%%%%%%%%%%%%
%%%%%%%%%%%%

The problem of the wave-function node was recognized 
in the original studies \cite{MN2,MN3}.
It was discussed that the robustness against the potential variation holds 
if the phase $\phi(t)$ is not divergent.
The divergence of the control Hamiltonian is also seen 
in the transitionless quantum driving.
In that case, the divergence is due to the level crossing 
and leads to a serious problem.
As for the present case, the divergence is considered to be 
a fictitious singularity 
which is not directly connected to any physical disaster.
It appears when we try to represent the potential by the $\sigma_z$ operator.
The original Hamiltonian $\hat{H}_{\rm FF}(t)$ does not include any 
singularity.

\begin{center}
\begin{figure}[t]
\begin{center}
\includegraphics[width=0.9\columnwidth]{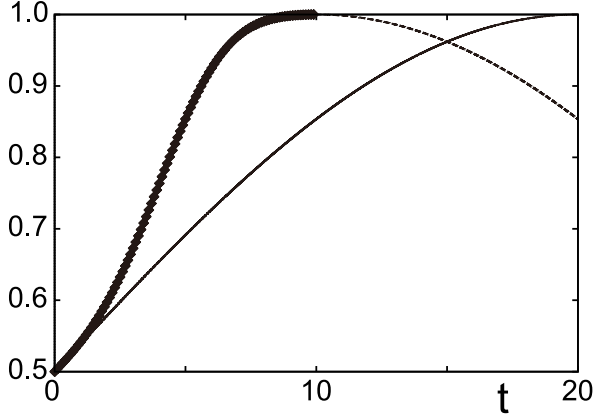}
\end{center}
\caption{
Probability of the up-spin state $|\langle{\sigma=1}|\psi(t)\rangle|^2$.
The solid line represents the result without the fast-forward scaling.
The bold and dashed lines represent 
the numerical and analytical results with the scaling, respectively.
$\bar{\alpha}=2.0$; $t_0=10$.
}
\label{fig3}
\end{figure}
\end{center}

Figure~\ref{fig3} shows the numerical result of 
the probability of the up-spin state 
$|\langle\sigma=+1|\psi_{\rm FF}(t)\rangle|^2$.
We can reach 
the final state at $t=20$ before the scaling  
in a shorter time $t=10$.
The numerical result agrees with the analytical one very well.

%%%%%%%%%%%%%%
%%%%%%%%%%%%

In the above example, 
Eq.~(\ref{condex}) is solved safely to find a real $\phi(t)$.
Generally, the condition (\ref{cond1}) does not always have a solution.
This is understood from the following example.
We consider the magnetic field 
\be
 \bm{h}(t)=\bmat{c}
 h(t)\cos\omega t \\ h(t)\sin\omega t \\ \omega
 \emat.
\ee
When we take the down-spin state $|-\rangle$ 
as the initial one $|\psi(0)\rangle$, 
the time dependence of the state is given by 
\be
 |\psi(t)\rangle = \bmat{c}
 -i\e^{-i\omega t/2}\sin\left(\frac{1}{2}\int_0^t\diff t'\,h(t')\right)  \\
 \e^{i\omega t/2}\cos\left(\frac{1}{2}\int^t_0\diff t'\,h(t')\right) 
 \emat.
\ee
The condition (\ref{cond1}) is written as 
\be
 \alpha(t)h(\Lambda(t)) = h(t)\cos\left(\phi(t)+\omega\Lambda(t)-\omega t\right).
\ee
Since $\alpha(t)\ge 1$, the solution can be found 
only when $h(t)$ is a decreasing function.

%%%%%%%%%%%%%%%%%%%%%%%%%%%%%%%%%%%%%%%%%%%%%%%%%%%%%%%%%%%%%%%%%%%%%%%%%%%
\subsection{Generalization}

It is straightforward to extend the method to systems 
in an $N$-dimensional Hilbert space.
In this case, we can choose the unitary transformation
\be
 \hat{U}(t) = \exp\left(-i\sum_{a=1}^{N-1}\phi_a(t)\hat{X}_a\right).
 \label{unitary}
\ee
The operators $\{\hat{X}_a\}_{a=1,2,\cdots,N-1}$ are traceless and 
commute with each other:
\be
 [\hat{X}_a,\hat{X}_b] = 0. \label{x}
\ee
In the $N$-dimensional Hilbert space, there exist $N-1$ independent 
diagonal traceless matrices.
For $N=2$, we have only one operator $\hat{X}=\sigma_z/2$ 
as we have already discussed.

In the $N$-dimensional case, the acceleration potential takes the form 
\be
 \hat{V}(t)=\frac{1}{N}v_0(t)+\sum_{a=1}^{N-1}v_a(t)\hat{X}_a.
 \label{V}
\ee
The first term is proportional to the identity operator.
We choose the coefficients $\{v_{0}(t),v_1(t),\cdots,v_{N-1}(t)\}$
so that the condition $\hat{H}_{\rm FF}(t)\sim \hat{H}(t)+\hat{V}(t)$ holds.

%%%%%%%%%%%%%%%%%%%%%%%%%%%%%%%%%%%%%%%%%%%%%%%%%%%%%%%%%%%%%%%%%%%%%%%%%%%
\subsection{Example: two-spin system}
\label{ex2}

To see how the generalization described above works well, 
we consider the second example in a two-spin system  
where the Hilbert space has four dimensions.
We consider the Hamiltonian
\be
 \hat{H}(t) &=& \sigma_z^{(1)}\sigma_z^{(2)}\sin\omega t
 -\frac{1}{2}(\sigma_x^{(1)}+\sigma_x^{(2)})\cos\omega t \no\\
 & & +\frac{i\omega}{4}(\sigma_y^{(1)}\sigma_z^{(2)}
 +\sigma_z^{(1)}\sigma_y^{(2)}),
\ee
where $\bm{\sigma}^{(1,2)}$ denote the Pauli matrices
for spins 1 and 2, respectively.
This is an example used in Ref.~\cite{OM} 
to study the transitionless driving.
The last term of the Hamiltonian corresponds to the counterdiabatic term.
One of the solutions of the Schr\"odinger equation is written as 
\be
 |\psi(t)\rangle = \frac{\e^{it}}{2\sqrt{1+\sin\omega t}}\bmat{c}
 \cos\omega t \\
 1+\sin\omega t \\
 1+\sin\omega t \\
 \cos\omega t \emat,
\ee
where we take the basis 
$\{ \left|\uparrow\uparrow\right>, \left|\uparrow\downarrow\right>, 
 \left|\downarrow\uparrow\right>, \left|\downarrow\downarrow\right>\}$ 
using the notations
$\sigma_z\left|\uparrow\right>=\left|\uparrow\right>$ and 
$\sigma_z\left|\downarrow\right>=-\left|\uparrow\right>$.
This state makes the transition from 
$|\psi(0)\rangle = \left|\rightarrow\rightarrow\right>$, 
both spins pointing in the $x$ direction, 
to an entangled state 
$|\psi(t_{\rm f}=\pi/2\omega)\rangle = 
(\left|\uparrow\downarrow\right>+\left|\downarrow\uparrow\right>)/\sqrt{2}$.

\begin{center}
\begin{figure}[t]
\begin{center}
\includegraphics[width=0.9\columnwidth]{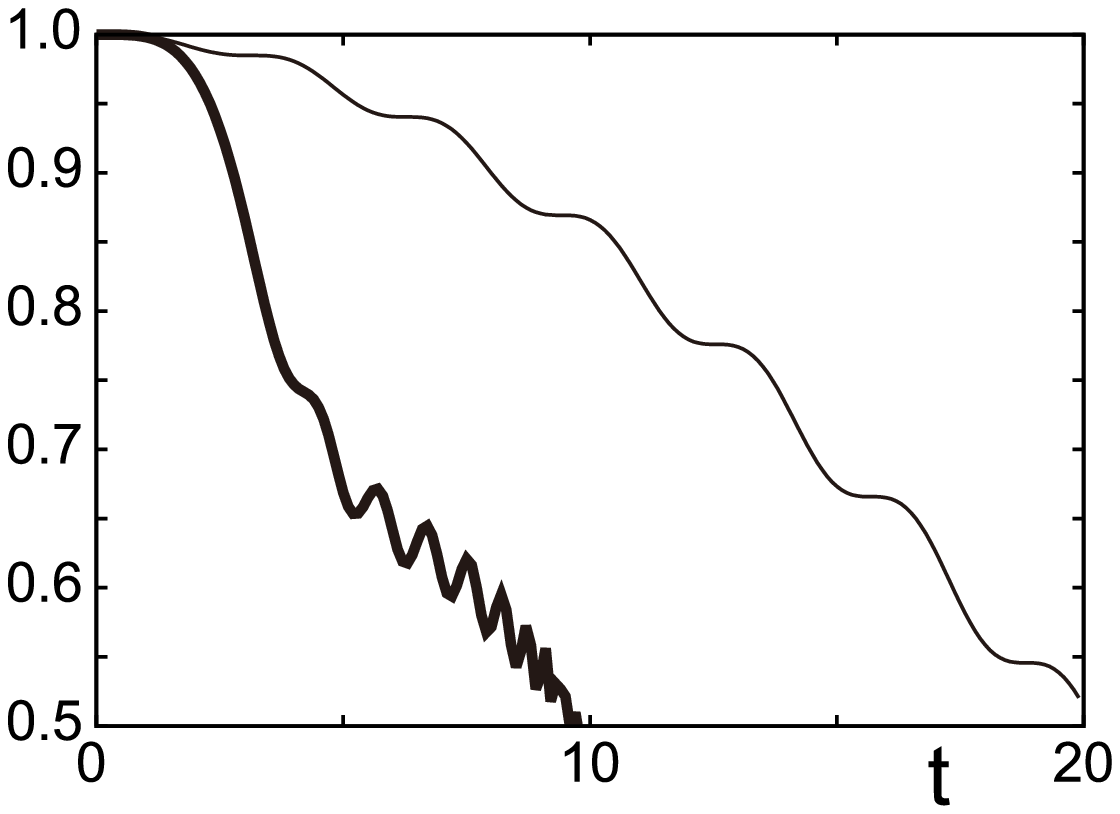}
\includegraphics[width=0.9\columnwidth]{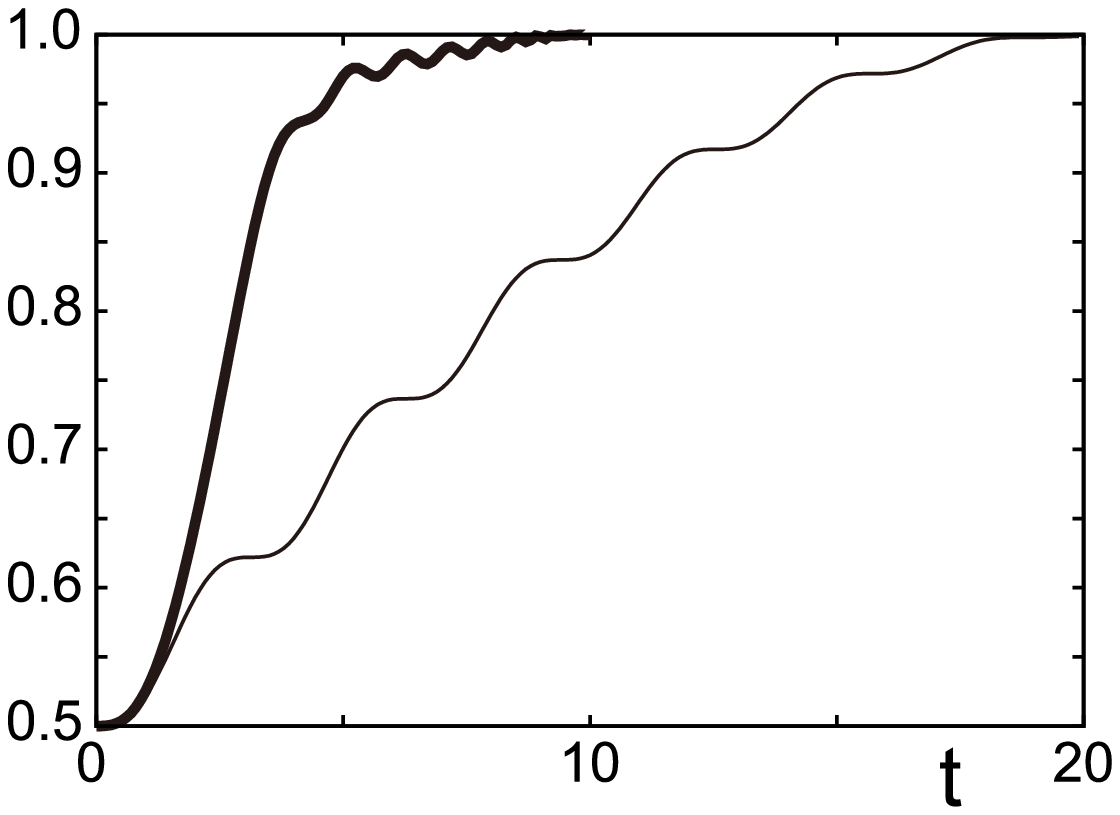}
\end{center}
\caption{
Overlap of the state with the initial state 
$|\langle\psi(0)|\psi(t)\rangle|^2$ (upper panel)
and with the final state 
$|\langle\psi(t_{\rm f})|\psi(t)\rangle|^2$ (lower).
The solid line represents the result without the fast-forward scaling 
and the bold with the scaling.
$\bar{\alpha}=2.0$; $t_0=10$.
}
\label{fig4}
\end{figure}
\end{center}

We use the unitary transformation (\ref{unitary2}) with 
\be
 \hat{f}(t)=\phi_1(t)\sigma_z^{(1)}+\phi_2(t)\sigma_z^{(2)}
 +\phi_3(t)\sigma_z^{(1)}\sigma_z^{(2)}.
\ee
Correspondingly, the acceleration potential takes the form
\be
 \hat{V}(t) &=& v_0(t)+v_1(t)\sigma_z^{(1)}
 +v_2(t)\sigma_z^{(2)} 
 +v_3(t)\sigma_z^{(1)}\sigma_z^{(2)}. \no\\
\ee
We use the same protocol as the example in Sec.~\ref{ex}.
Following the same manipulation, 
we obtain the conditions for $\phi_{1,2,3}(t)$ as 
\be
 && \phi_1(t)=\phi_2(t)=0, \\
 && 2\cos\omega t\sin 2\phi_3(t)= \alpha(t)\omega.
\ee
Using $\phi_3(t)$, we can write the potential as 
\be
 && v_0(t)=-1+\frac{\cos\omega t\cos 2\phi_3(t)}{\cos\omega\Lambda(t)}, \\
 && v_1(t)=v_2(t)=0, \\
 && v_3(t)= \dot{\phi}_3(t)+\alpha(t)\sin\omega\Lambda(t)
 -\sin\omega t \no\\
 && -\sin\omega\Lambda(t)
 +\tan\omega\Lambda(t)\cos\omega t\cos 2\phi_3(t).
\ee
In the present example, the system can be accelerated 
by controlling the exchange interaction in a proper way
as we show the numerical calculation in Fig.~\ref{fig4}.
%%%%%%%%%%%%%%
%%%%%%%%%%%%

%%%%%%%%%%%%%%%%%%%%%%%%%%%%%%%%%%%%%%%%%%%%%%%%%%%%%%%%%%%%%%%%%%%%%%%%%%%
\section{Acceleration of adiabatic states}

%%%%%%%%%%%%%%%%%%%%%%%%%%%%%%%%%%%%%%%%%%%%%%%%%%%%%%%%%%%%%%%%%%%%%%%%%%%
\subsection{Transitionless quantum driving}

We have demonstrated that the fast-forward scaling can be applied 
to systems in a finite-dimensional space.
In this section, we consider the acceleration of adiabatic states.  
In the transitionless quantum driving, the Hamiltonian consists of 
the adiabatic and counterdiabatic terms: 
$\hat{H}(t)=\hat{H}_{\rm ad}(t)+\hat{H}_{\rm cd}(t)$.
Each term is given respectively by 
\be
 && \hat{H}_{\rm ad}(t) = \sum_{n}E_n(t)|n(t)\rangle\langle n(t)|, \label{ad} \\
 && \hat{H}_{\rm cd}(t) = i\sum_{m\ne n}|m(t)\rangle
 \langle m(t)|\dot{n}(t)\rangle\langle n(t)|, \label{cd}
\ee
where $|n(t)\rangle$ is the eigenstate of $\hat{H}_{\rm ad}(t)$ 
with the eigenvalue $E_n(t)$.
The eigenstates are orthonormalized and satisfy the completeness relation.
It can be shown that the solution of the Schr\"odinger equation is given by 
the adiabatic state of $\hat{H}_{\rm ad}(t)$~\cite{DR1,DR2,Berry}.

We apply the fast-forward scaling to the transitionless driving.
To achieve this, we first use the unitary transformation (\ref{unitary}).
Then, the fast-forward Hamiltonian is written as
\be
 \hat{H}_{\rm FF}(t) = \sum_a\dot{\phi}_a(t)\hat{X}_a
 +\hat{U}(t)\hat{H}(\Lambda(t))\hat{U}^\dag(t). \label{Hff}
\ee
Using the new basis 
\be
 |\tilde{n}(t)\rangle = \hat{U}(t)|n(\Lambda(t))\rangle,
\ee
we can write 
$\hat{H}_{\rm FF}(t)=\hat{H}_{\rm FF}^{\rm ad}(t)+\hat{H}_{\rm FF}^{\rm cd}(t)$ with 
\be
 \hat{H}^{\rm ad}_{\rm FF}(t) &=& \sum_{n}\left(\alpha(t)E_n(\Lambda(t))
 +\sum_a\dot{\phi}_a(t)\langle\tilde{n}(t)|\hat{X}_a|\tilde{n}(t)\rangle\right)
 \no\\
 &&\times|\tilde{n}(t)\rangle\langle\tilde{n}(t)|, \\
 \hat{H}^{\rm cd}_{\rm FF}(t) &=& 
 i\sum_{m\ne n}|\tilde{m}(t)\rangle\langle
 \tilde{m}(t)|\dot{\tilde{n}}(t)\rangle\langle \tilde{n}(t)|.
\ee
This expression denotes that the fast-forward state of 
the transitionless driving is also transitionless.
We note that the first term of Eq.~(\ref{Hff}) 
does not affect the counterdiabatic part.

Next, we determine the fast-forward potential $\hat{V}(t)$ 
such that the relation 
$\hat{H}_{\rm FF}(t)\sim\hat{H}(t)+\hat{V}(t)$
holds.
We consider the case where the state is in a specific eigenstate 
$|\tilde{n}(t)\rangle$.
We impose the condition 
\be
 \hat{H}_{\rm FF}(t)|\tilde{n}(t)\rangle 
 = (\hat{H}(t)+\hat{V}(t))|\tilde{n}(t)\rangle 
\ee
to find the potential in a diagonal form (\ref{V}).
The coefficients $\{v_0,v_1,\ldots,v_{N-1}\}$ are determined 
from 
\be
 &&\langle\tilde{n}|\hat{H}(t)|\tilde{n}\rangle
 = \alpha E_{n\Lambda}-\frac{1}{N}v_0+\sum_a(\dot{\phi}_a-v_a)
 \langle\tilde{n}|\hat{X}_a|\tilde{n}\rangle, \no\\ \label{condad1}\\
 &&\langle\tilde{m}|\hat{H}(t)|\tilde{n}\rangle
 = i\langle\tilde{m}|\dot{\tilde{n}}\rangle
 -\sum_av_a\langle\tilde{m}|\hat{X}_a|\tilde{n}\rangle, \label{condad2}
\ee
where $E_{n\Lambda}=E_n(\Lambda(t))$ and 
$m$ takes values different from $n$.
There are $N$-independent equations.
The equations imply that the solution depends on
the state $n$ to use.

We examine the two-level case where $\hat{X}=\sigma_z/2$.
The transitionless driving is achieved by the Hamiltonian 
\be
 \hat{H}(t) &=& \frac{1}{2}\left(
 \bm{h}(t)+\frac{\bm{h}(t)\times\dot{\bm{h}}(t)}{\bm{h}^2(t)}\right)
 \cdot\bm{\sigma},
\ee
where the second term denotes the counterdiabatic field~\cite{DR1, DR2, Berry}.
Using the polar coordinate representation of the magnetic field
\be
 \bm{h}(t)=h(t)\bmat{c} 
 \sin\theta(t)\cos\varphi(t) \\
 \sin\theta(t)\sin\varphi(t) \\ \cos\theta(t) \emat,
\ee
we can write the eigenstates of 
$\hat{H}_{\rm ad}(t)=\bm{h}(t)\cdot\bm{\sigma}/2$ as
\be
 |n(t)\rangle = \left\{
 \bmat{c} \cos\frac{\theta(t)}{2} \\ 
 \e^{i\varphi(t)}\sin\frac{\theta(t)}{2} \emat, \ 
 \bmat{c} -\e^{-i\varphi(t)}\sin\frac{\theta(t)}{2} \\ 
 \cos\frac{\theta(t)}{2} \emat\right\}. \no\\ \label{n}
\ee
We consider an acceleration of the former state 
under the choice of parameters 
\be
 && h(t)=1, \\
 && \theta(t)=\frac{\pi}{2}-\omega t, \\
 && \varphi(t)=0.
\ee
This is the example treated in Sec.~\ref{ex}.
Equations (\ref{condad1}) and (\ref{condad2}) are written as 
\be
 & & \alpha-v_0+(\dot{\phi}-v)\sin\omega\Lambda  \no\\
 &=& \sin\omega t\sin\omega\Lambda +\cos\omega t\cos\omega\Lambda\cos\phi
 -\omega\cos\omega\Lambda\sin\phi, \no\\ \\
 & & i\alpha\omega+(\dot{\phi}-v)\cos\omega\Lambda \no\\
 &=& \sin\omega t\cos\omega\Lambda-\cos\omega t\sin\omega\Lambda\cos\phi
 \no\\
 &&  +i\cos\omega t\sin\phi
 +i\omega\cos\phi
 +\omega\sin\omega\Lambda\sin\phi,
\ee
which gives the condition (\ref{condex}) and the potential (\ref{Vs}).

Thus we find that the fast-forward scaling is useful when accelerating 
the adiabatic state.
We note that the original fast-forward Hamiltonian 
$\hat{H}_{\rm FF}(t)=\hat{H}_{\rm FF}^{\rm ad}(t)+\hat{H}_{\rm FF}^{\rm cd}(t)$
in Eq.~(\ref{Hff})
is enough to accelerate the state evolution.
The advantage of the fast-forward scaling is that 
we can take the form of the acceleration potential in a diagonal form.
This arbitrariness of the counterdiabatic Hamiltonian 
comes from the fact that the acceleration potential 
depends on the state to accelerate.
The counterdiabatic Hamiltonian (\ref{cd})
is applied to any states $\{|n(t)\rangle\}$.
If we consider a specific state $|n(t)\rangle$ only, 
it is possible to deform the counterdiabatic Hamiltonian as 
\be
 \hat{H}_{\rm cd}^{(n)}(t) &=& i(1-|n(t)\rangle\langle n(t)|)
 |\dot{n}(t)\rangle\langle n(t)|+({\rm H.c.}) \no\\
 && +(\mbox{$n$-independent terms}),
\ee
where $n$-independent terms can be taken arbitrarily.
This property was used in Refs.~\cite{dCRZ,Takahashi2} 
to apply the method in many-body systems
where the counterdiabatic Hamiltonian (\ref{cd}) 
takes a complicated form.
In Ref.~\cite{OM}, 
the same idea was used to obtain the partial suppression of 
the nonadiabatic transitions.
In the present case, using the arbitrariness of 
the counterdiabatic Hamiltonian,
we choose the acceleration potential in a diagonal form.

%%%%%%%%%%%%%%%%%%%%%%%%%%%%%%%%%%%%%%%%%%%%%%%%%%%%%%%%%%%%%%%%%%%%%%%%%%%
\subsection{Lewis-Riesenfeld invariant}

It is also possible to formulate the fast-forward scaling
by using the invariant-based engineering.
We consider the operator $\hat{F}(t)$ satisfying 
\be
 i\frac{\partial\hat{F}(t)}{\partial t}=[\hat{H}(t),\hat{F}(t)].
 \label{LReq}
\ee
This operator is called the Lewis-Riesenfeld invariant~\cite{LR}
and the eigenvalues $\lambda_n$ are independent of time: 
\be
 \hat{F}(t)=\sum_n\lambda_n|n(t)\rangle\langle n(t)|, \label{LR}
\ee
where $|n(t)\rangle$ represents the corresponding eigenstate.
In the invariant-based engineering, we construct the Hamiltonian 
for a given invariant~\cite{CRSCGM}.
Using the basis of the eigenstates of $\hat{F}(t)$, 
we can write the Hamiltonian as 
$\hat{H}(t)=\hat{H}_{\rm ad}(t)+\hat{H}_{\rm cd}(t)$, where 
each term is given by Eqs.~(\ref{ad}) and  (\ref{cd}), respectively.
Thus the time evolution becomes transitionless.
For example, in the case of the Hamiltonian in Sec.~\ref{ex}, 
the invariant is given by
\be
 \hat{F}(t) &=& \frac{\lambda_+}{2}\bmat{cc} 1+\sin\omega t & \cos\omega t \\ 
 \cos\omega t & 1-\sin\omega t \emat \no\\
 && +\frac{\lambda_-}{2}\bmat{cc} 1-\sin\omega t & -\cos\omega t \\ 
 -\cos\omega t & 1+\sin\omega t \emat.
\ee
This is obtained by substituting the form of the eigenstates in Eq.~(\ref{n}) 
to Eq.~(\ref{LR}).

We apply the fast-forward scaling to the equation 
for the invariant (\ref{LReq}).
First, using the scaling $\Lambda(t)$ and 
the unitary transformation (\ref{unitary}), 
we obtain 
\be
 i\frac{\partial\hat{F}_{\rm FF}(t)}{\partial t}
 = [\hat{H}_{\rm FF}(t),\hat{F}_{\rm FF}(t)],
\ee
where $\hat{H}_{\rm FF}(t)$ is given in Eq.~(\ref{Hff}) and 
\be
 \hat{F}_{\rm FF}(t) = \hat{U}(t)\hat{F}(\Lambda(t))\hat{U}^\dag(t) 
 = \sum_n\lambda_n|\tilde{n}(t)\rangle\langle\tilde{n}(t)|. 
\ee
Second, we determine the acceleration potential $\hat{V}(t)$ in 
the form of Eq.~(\ref{V}) satisfying  
\be
 i\frac{\partial\hat{F}_{\rm FF}(t)}{\partial t}
 =[\hat{H}(t)+\hat{V}(t),\hat{F}_{\rm FF}(t)]. \label{LRFFeq}
\ee
As we explained in the previous subsection, 
the solution depends on the state to accelerate.
In the present case, the choice of the state is reflected in 
the eigenvalues $\{\lambda_n\}$.
Here we consider the simplest case where one of the eigenvalues is one 
and the others are zero:
\be
 \hat{F}(t)=|n(t)\rangle\langle n(t)|.
\ee
From Eq.~(\ref{LRFFeq}), we have 
\be
 i\langle\tilde{m}|\dot{\tilde{n}}\rangle 
 = \langle\tilde{m}|\hat{H}(t)|\tilde{n}\rangle
 +\sum_av_a\langle\tilde{m}|\hat{X}_a|\tilde{n}\rangle,
\ee
where $m$ takes values different from $n$.
This equation coincides with Eq.~(\ref{condad2}).
The diagonal part of the Hamiltonian cannot be determined from 
Eq.~(\ref{LRFFeq}) since it does not contribute to the equation.
We impose 
\be
 \langle\tilde{n}|(\hat{H}(t)+\hat{V}(t))|\tilde{n}\rangle
 = \langle\tilde{n}|\hat{H}_{\rm FF}(t)|\tilde{n}\rangle.
\ee
This is equivalent to Eq.~(\ref{condad1}).
Thus, in the present formulation based on the invariant, 
we find the same result as the formulation using the transitionless driving.
The derivation denotes that 
the form of the acceleration potential 
depends on the choice of the eigenvalues of the invariant.
It corresponds to setting the initial condition for the time evolution.

%%%%%%%%%%%%%%%%%%%%%%%%%%%%%%%%%%%%%%%%%%%%%%%%%%%%%%%%%%%%%%%%%%%%%%%%%%%
\section{Summary}

In summary, we have shown that the method of the fast-forward scaling 
is applicable to systems in a finite-dimensional Hilbert space.
The unitary transformation (\ref{unitary}) is utilized 
to have the acceleration potential in a diagonal form.

Although the use of the fast-forward scaling is not restricted 
to the acceleration of the adiabatic state, 
we find that the method is most useful when it is applied 
to the transitionless driving.
In that case, the fast-forward state follows 
a different adiabatic passage and 
is understood as a different transitionless driving.
The advantage of using the transitionless driving is that 
the general condition to determine the acceleration potential 
can be explicitly written as (\ref{condad1}) and (\ref{condad2}).
Using both methods together, we can consider the efficient 
acceleration of the state.

The form of the acceleration potential is state dependent. 
Different states require different auxiliary 
fast-forward driving potentials.
This limitation is absent in the transitionless driving 
which applies to an arbitrary state. 
Furthermore, our analysis implies that 
the acceleration is not always possible.
The condition of determining the unitary transformation 
sometimes fails to find the acceleration potential.

As a related problem, it will be interesting to know 
the relation of the fast-forward scaling to the method proposed 
in Ref.~\cite{delCampo}. 
Using the framework of transitionless quantum driving, 
the counterdiabatic driving was derived for a large family of 
many-body and nonlinear systems under scale-invariant dynamics. 
It may be interesting to study the fast-forward scaling 
in such systems, and to explore the prospects of extending 
the fast-forward technique beyond
single-particle and mean-field descriptions.

%%%%%%%%%%%%%%%%%%%%%%%%%%%%%%%%%%%%%%%%%%%%%%%%%%%%%%%%%%%%%%%%%%%%%%%%%%%%
\section*{Acknowledgments}

The author is grateful to X.~Chen, S.~Masuda, J.~G.~Muga, and M.~Nakahara 
for stimulating discussions. 

%%%%%%%%%%%%%%%%%%%%%%%%%%%%%%%%%%%%%%%%%%%%%%%%%%%%%%%%%%%%%%%%%%%%%%%%%%%%
%%%%%%%%%%%%%%%%%%%%%%%%%%%%%%%%%%%%%%%%%%%%%%%%%%%%%%%%%%%%%%%%%%%%%%%%%%%%

\end{document}